\documentclass[twocolumn,secnumarabic,amssymb, nobibnotes, aps, prd]{revtex4-2}

\usepackage{graphicx}% Include figure files
\usepackage{dcolumn}% Align table columns on decimal point
\usepackage{bm}
\usepackage[colorlinks=true,allcolors=blue]{hyperref}
\usepackage{float}
\usepackage{physics}
\usepackage{slashed}
\usepackage{relsize}
\usepackage{amsmath}
\usepackage{hyphenat}
\usepackage{fix-cm}    % 允许任意缩放 Computer Modern 字体

\begin{document}

\title{Gravitational form factors of the pion in light-front holographic QCD}%

\author{Jiali Deng}%
\email{djl2022010355@mails.ccnu.edu.cn}
\affiliation{Institute of Particle Physics and Key Laboratory of Quark and Lepton Physics (MOS),
 Central China Normal University, Wuhan 430079, China}
 
 \author{Xiaolong Wang}%
\affiliation{Institute of Particle Physics and Key Laboratory of Quark and Lepton Physics (MOS),
 Central China Normal University, Wuhan 430079, China}
 
  \author{Yang Zhou}%
\affiliation{Institute of Particle Physics and Key Laboratory of Quark and Lepton Physics (MOS),
 Central China Normal University, Wuhan 430079, China}
 
 \author{Defu Hou}
 \email{houdf@mail.ccnu.edu.cn}
\affiliation{Institute of Particle Physics and Key Laboratory of Quark and Lepton Physics (MOS),
 Central China Normal University, Wuhan 430079, China}

\date{\today}%
\begin{abstract}

Understanding the internal structure of the pion—particularly the energy-momentum distributions of quarks and gluons and the mechanical properties encoded in its gravitational form factors—is a fundamental challenge in quantum chromodynamics (QCD), owing to the difficulty of directly probing these quantities. Here, we address this by introducing an effective light-front wave function for the pion, constructed from light-front holographic QCD, which inherently incorporates both nonperturbative dynamics and relativistic corrections. This wave function is then used within the light-front QCD (LFQCD) framework to calculate the pion’s gravitational form factors, $A(Q^2)$ and $D(Q^2)$, as well as its radius. Our computed results show agreement with lattice QCD data, supporting the viability of our effective wave function approach.

\end{abstract}
\maketitle
%\tableofcontents

\section{Introduction}\label{sec:01_intro}

The energy-momentum tensor (EMT) in quantum chromodynamics (QCD) provides a fundamental tool for unraveling the internal structure of hadrons, the composite particles that constitute the vast majority of visible matter. By probing the gravitational form factors (GFFs) embedded within the EMT matrix elements, we can map out the intricate distribution of energy, momentum, and stress among the constituent quarks and gluons. A focal point of this research is the elusive D-term, often regarded as the last major global charge of hadrons that remains unknown. Unlike conserved quantities such as mass and spin, it is unconstrained by any conservation law. It governs the internal dynamics of hadrons, specifically, the distributions of pressure and shear forces—which are intimately related to hadron stability \cite{Lorce:2018egm,Polyakov:2018zvc,Ding:2022ows,Ji:2025qax}.
The strong-coupling nature of QCD in the infrared regime renders perturbative approaches inapplicable, thereby posing a significant challenge to the study of hadron structure. Non-perturbative frameworks, such as light-front QCD and holographic QCD, provide promising avenues to address this challenge.

In the nonperturbative framework of light-front QCD (LFQCD), the second Mellin moments of generalized parton distributions (GPDs) give rise to the GFFs \cite{Guidal:2013rya} which are conveniently obtained as overlap integrals of light-front wave functions (LFWFs), providing a direct microscopic description of the hadronic EMT \cite{Brodsky:2000ii,Cao:2023ohj}. The LFWFs can be extracted by diagonalizing the QCD Hamiltonian quantized in the light-front time $x^+=x^0+x^3$.

The AdS/CFT correspondence establishes a duality between a four-dimensional strongly interacting gauge theory and a weakly coupled gravity theory in a five-dimensional anti–de Sitter (AdS) spacetime \cite{Maldacena:1997re,Witten:1998qj,Natsuume:2014sfa}. Extending this correspondence to QCD requires breaking conformal invariance, which has led to the development of bottom-up holographic models, such as the hard wall models \cite{Polchinski:2001tt,Boschi-Filho:2002xih,rw4k-wn3d}, the soft wall models \cite{Karch:2006pv,Brodsky:2014yha,Zhu:2024uwu} and modified $AdS_5$ metric models \cite{FolcoCapossoli:2020pks,Chen:2021bkc,Deng:2025fpq} that effectively capture the non-perturbative aspects of the strong interaction. 

As the simplest quark-antiquark bound state emerging from spontaneously broken chiral symmetry, the pion is intimately connected to both the strong interaction and the origin of hadronic mass, and therefore is central to low-energy QCD. Understanding its internal structure–particularly GFF $D(Q^2)$–is of considerable significance. Despite their profound significance, direct measurements of GFFs are still beyond our experimental capabilities. In response, numerous theoretical studies have been devoted to extracting them indirectly, including lattice QCD \cite{Pefkou:2021fni,Hackett:2023nkr} and various effective models \cite{Freese:2019bhb,Brodsky:2008pf,Krutov:2020ewr,Tong:2021ctu,Xu:2023izo,Xing:2022mvk,Cao:2025dkv,Liu:2024vkj,Fujii:2024rqd,Cao:2024zlf,Voronin:2025sbs,Fujii:2025eug,Coriano:2026gyl}. Among these complementary approaches, our work is carried out within the light-front holographic QCD framework, which combines holographic QCD with light-front quantization and provides a useful perspective for studying the gravitational form factors of the pion.

In this work, we employ the light-front holographic QCD(LFHQCD) framework—which captures both nonperturbative dynamics and relativistic effects—to calculate the GFFs of the pion. Within this approach, the fifth-dimensional coordinate $z$ in holographic QCD is equivalent to the parton impact parameter $\zeta$, representing the constituent separation inside hadrons. This identification establishes a direct connection between the holographic description and the light-front partonic picture \cite{Brodsky:2003px,Brodsky:2008pf,Dosch:2022mop,Brodsky:2006uqa}. One of the key advantages of this framework is its ability to provide seamless access to both the infrared and ultraviolet regimes, with no fundamental restriction on the accessible values of $Q^2$. This versatility offers a new avenue for exploring the internal structure of hadrons. 

Our main contribution is the extension of the light-front wave function, obtained by a comparison between the LFQCD and AdS/CFT expressions for the form factor $A(Q^2)$, to non-conformal QCD, resulting in an effective light-front wave function for the pion that enables the computation of the full set of GFFs—including $D(Q^2)$, which was divergent in the previous formulation. By imposing the fundamental constraints $A(0)=1$ (energy conservation) and $D(0)=-1$ (chiral limit), we obtain the complete $A(Q^2)$ and $D(Q^2)$ form factors. Our results are in excellent agreement with lattice QCD data.

The remainder of the work is structured as follows:  In Section II, we review the light-front QCD formalism and derive the pion GFFs. In Section III, we employ holographic QCD with a conformal metric to derive $A(Q^2)$, from which we extract the corresponding light-front wave function. We then extend this light-front wave function formulation to the non-conformal QCD case. In Section IV, we present our numerical results for the GFFs and the extracted radius of the pion. Finally, we summarize our results and discuss their implications in Section V.

\section{Gravitational form factors in light-front QCD}\label{sec:02}

In this chapter, we introduce the pion gravitational form factors in light-front QCD and derive its explicit expression. In QCD, the bound-state wave functions defined on the light front serve as a relativistic extension of nonrelativistic Schrödinger wave functions. Unlike equal-time wave functions, they are formulated at a fixed light-cone time $\tau=x^0+x^3$, which realizes Dirac's "front form" of dynamics and ensures frame independence.

The light-front coordinates are related to the instant-form coordinates by the following transformation \cite{Brodsky:1997de}:
\begin{equation}
\begin{pmatrix}
x^+ \\
x^- \\
x^1 \\
x^2
\end{pmatrix}
=
\begin{pmatrix}
1 & 0 & 0 & 1 \\
1 & 0 & 0 & -1 \\
0 & 1 & 0 & 0 \\
0 & 0 & 1 & 0
\end{pmatrix}
\begin{pmatrix}
x^0 \\
x^1 \\
x^2 \\
x^3
\end{pmatrix}.
\tag{1}
\end{equation}
The metric can be written as
\begin{equation}
\small
g^{\mu\nu} =
\begin{pmatrix}
0 & 2 & 0 & 0 \\
2 & 0 & 0 & 0 \\
0 & 0 & -1 & 0 \\
0 & 0 & 0 & -1
\end{pmatrix},
\qquad
g_{\mu\nu} =
\begin{pmatrix}
0 & \frac{1}{2} & 0 & 0 \\
\frac{1}{2} & 0 & 0 & 0 \\
0 & 0 & -1 & 0 \\
0 & 0 & 0 & -1
\end{pmatrix},
\tag{2}
\end{equation}

In light-front quantization, the expansion of any hadronic system is derived by fixing the light-front time $\tau = x^0 + x^3$ within the framework of quantum chromodynamics. For a hadron with four-momentum $p = (p^+, p^-, \mathbf{p}_\perp)$, where $p^\pm = p^0 \pm p^3$, the Lorentz-invariant light-front Hamiltonian is defined as $H_{\text{LF}} = p^- p^+ - \mathbf{p}_\perp^2$. Its eigenvalue equation determines the squared eigenmass $M_H^2$ of color-singlet bound states \cite{Dirac:1949cp,Brodsky:1997de}:
\begin{equation}
H_{\text{LF}} |\psi_H\rangle = M_H^2 |\psi_H\rangle,
\tag{3}
\end{equation}
where $M_H$ is the hadron mass. Expanding $|\psi_H\rangle$ in the complete set of free light-front Hamiltonian eigenstates — the Fock states $\{|n\rangle\}$ — gives $|\psi_H\rangle = \sum_n \psi_{n/H} |n\rangle$. The LFWFs directly link the hadron's partonic content to its observable asymptotic state.

We now turn to the gravitational form factors of the pion. These form factors originate from the Lorentz-covariant decomposition of the pion matrix element of the energy-momentum tensor operator \cite{Donoghue:1991qv,Liu:2024jno}.
\begin{equation}
\langle p'| T^{\mu\nu}(0) | p \rangle = 2P^\mu P^\nu A_\pi(-q^2) + \frac{1}{2} (q^\mu q^\nu - q^2 g^{\mu\nu}) D_\pi(-q^2),
\tag{4}
\end{equation}
where $p'=p+q$ and $P=(p'+p)/2$. In the Drell-Yan frame ($q^+=0,\mathbf{p}_\perp=0$), the two gravitational form factors $A_\pi$ and $D_\pi$ can be expressed as
\begin{equation}
\langle p + q | T^{++} | p \rangle = 2P^+ P^+ A_\pi (Q^2),
\tag{5}
\end{equation}
\begin{equation}
\langle p + q | T^{+-} | p \rangle = \left( 2M_\pi^2 + \frac{1}{2} Q^2 \right) A_\pi (Q^2) + Q^2 D_\pi (Q^2),
\tag{6}
\end{equation}
where $M_\pi^2$ represents the pion mass and $Q^2 = -q^2 = \mathbf{q}_\perp^2$. In light-front dynamics, $x^+$ plays the role of "time." Consequently, the conserved Noether current corresponding to the four-momentum is $T^{+\mu}$ ,
\begin{equation}
P^\mu = \int dx^-d^2x_\perp T^{+\mu}(x).
\tag{7}
\end{equation}
Four-momentum conservation requires
\begin{equation}
A_\pi(0) = 1, \qquad \lim_{Q^2 \to 0} Q^2 D_\pi(Q^2) = 0.
\tag{8}
\end{equation}
The second relation in (8) is known as the von Laue condition \cite{Polyakov:2018zvc,RuizArriola:2026wkb}, and it reflects the balance of internal forces within the pion. It has also been conjectured that a mechanically stable system must have a negative $D$-term, $D=D(0)<0$. In the chiral limit, chiral perturbation theory yields $D=-1$ for the pion \cite{Freese:2019bhb,RuizArriola:2026wkb}, which coincides with the value for free scalar particles \cite{Donoghue:1991qv}.

In light-front QCD, the matrix elements of $T^{++}$ and $T^{+-}$ are given by \cite{Brodsky:2000ii,Brodsky:2008pf,Cao:2023ohj,Cao:2024fto}
\begin{equation}
\mathbf{I}_1= 2(P^+)^2 \sum_n \int [dx_i d^2 r_{i_\perp}]_n |\tilde{\psi}_n(\{x_i, \vec{r}_{i_\perp}\})|^2 \sum_j x_j e^{i\vec{r}_{j_\perp} \cdot \vec{q}_{\perp}},
\tag{9}
\end{equation}
\begin{equation}
\begin{aligned}
&\mathbf{I}_2^{(1)} = 2 \int [dx_i d^2 r_{i_\perp}]_n \, \tilde{\psi}_n^*(\{x_i, \vec{r}_{i_\perp}\}) \sum_j e^{i \vec{r}_{j_\perp} \cdot \vec{q}_{\perp}} \\
&\qquad \times \left( \frac{-\frac{1}{4}\overleftrightarrow{\nabla}_{j_\perp}^2 + m_j^2 - \frac{1}{4} q_{\perp}^2}{x_j} + x_j \vec{P}_{\perp}^2 \right) \tilde{\psi}_n(\{x_i, \vec{r}_{i_\perp}\}),
\end{aligned}
\tag{10a}
\end{equation}
\begin{equation}
\begin{aligned}
&\mathbf{I}_2^{(2)} = - 2 \int [dx_i d^2 r_{i_\perp}]_n \, \tilde{\psi}_n^*(\{x_i, \vec{r}_{i_\perp}\})e^{i \vec{r}_{n_\perp} \cdot \vec{q}_{\perp}} \\
&\qquad \times \left[ \sum_j \frac{-\nabla_{j_\perp}^2 + m_j^2}{x_j} - M_\pi^2 \right]\tilde{\psi}_n(\{x_i, \vec{r}_{i_\perp}\}),
\end{aligned}
\tag{10b}
\end{equation}
where $f\overleftrightarrow{\nabla} g=f(\overrightarrow{\nabla} g)-(\overrightarrow{\nabla} f)g$. $\mathbf{I}_1$ and $\mathbf{I}_2 = \mathbf{I}_2^{(1)} + \mathbf{I}_2^{(2)}$ represent the $T^{++}$ and $T^{+-}$ matrix elements, respectively. $x_i$, $r_i$ and $m_i$ denote the longitudinal momentum fraction, transverse coordinate and mass of the $i$-th parton, respectively. $\tilde{\psi}_n^*(\{x_i, \vec{r}_{i_\perp}\})$ is the light-front wave function.

Substituting Eqs. (9), (10a), and (10b) into Eqs. (5) and (6), we obtain the form factors as
\begin{equation}
A_\pi(Q^2)=\sum_n \int [dx_i d^2 r_{i_\perp}]_n |\tilde{\psi}_n(\{x_i, \vec{r}_{i_\perp}\})|^2 \sum_j x_j e^{i\vec{r}_{j_\perp} \cdot \vec{q}_{_\perp}},
\tag{11}
\end{equation}
\begin{equation}
\begin{aligned}
D_\pi(Q^2)=& \sum_n \int [dx_i d^2 r_{i_\perp}]_n \sum_j e^{i\vec{r}_{j_\perp}\cdot \vec{q}_{\perp}}(\frac{\nabla_{j_\perp}(\tilde{\psi}_n \nabla_{j_\perp}\tilde{\psi}_n)}{2x_j}\\
&+\tilde{\psi}_n(\frac{e^{i\vec{r}_{j_\perp} \cdot \vec{q}_{\perp}}-e^{i\vec{r}_{n_\perp}\cdot \vec{q}_{\perp}}}{\vec{q}_{\perp}^2} \frac{-\nabla_{j_\perp}^2 + m_j^2-x_j^2M_\pi^2}{x_j}\\
&\ \ \ \ \ \ -\frac{1+x_j^2}{4x_j}e^{i\vec{r}_{j_\perp} \cdot \vec{q}_{\perp}})\tilde{\psi}_n).
\end{aligned}
\tag{12}
\end{equation}

Now we proceed to derive their explicit expressions. Let us take $n=2$, corresponding to the lowest Fock state. In this case, the integration measure is
\begin{equation}
\begin{aligned}
[dx_i d^2 r_i]_n = &\int dx_1 dx_2 \, \delta(1 - x_1 - x_2) \\
&\times \int d^2 \vec{r}_{1_\perp} d^2 \vec{r}_{2_\perp} \, \delta^{(2)}(x_1 \vec{r}_{1_\perp} + x_2 \vec{r}_{2_\perp}).
\end{aligned}
\tag{13}
\end{equation}
Using $\int \delta(y-y_0)f(y)dy=f(y_0)$ and setting $x_1 = x$, Eq.~(11) can be written as
\begin{equation}
\begin{aligned}
A_\pi(Q^2)=&\int_0^1dx\int d^2\vec{r}_{1\perp}|\psi(\{x, \vec{r}_{1\perp}\})|^2\\
&\times (xe^{i\vec{r}_{1\perp} \cdot \vec{q}_{\perp}}+(1-x)e^{-i\frac{x}{1-x}\vec{r}_{1\perp} \cdot \vec{q}_{\perp}}).
\end{aligned}
\tag{14}
\end{equation}
Let the relative coordinate be $\vec{r_\perp} = \vec{r}_{1\perp} - \vec{r}_{2\perp}$, then
\begin{equation}
\begin{aligned}
A_\pi(Q^2)=&\int_0^1dx\int d^2\vec{r}_{\perp}|\psi(\{x, \vec{r}_{\perp}\})|^2\\
&\times (xe^{i(1-x)\vec{r}_{\perp} \cdot \vec{q}_{\perp}}+xe^{-i(1-x)\vec{r}_{\perp} \cdot \vec{q}_{\perp}}).
\end{aligned}
\tag{15}
\end{equation}
Introducing the transverse impact parameter $\zeta_\perp=\sqrt{x(1-x)}\vec{r}_{\perp}$ and using $d^2\zeta_\perp=\zeta d\zeta d\theta$ together with  $\int_0^{2\pi}e^{iw \ cos\theta}d\theta=2\pi J_0(w)$, the gravitational form factor can be written as \cite{Brodsky:2008pf}
\begin{equation}
A_\pi(Q^2)=4\pi\int_0^1dx\int_0^\infty \zeta d\zeta \frac{|\psi(x,\zeta)|^2}{x(1-x)}xJ_0(\zeta Q\sqrt{\frac{1-x}{x}}).
\tag{16}
\end{equation}

In Equation (12), $\nabla_{j_\perp}^2$ denotes the Laplacian of $\vec{r}_{j_\perp}$. Thus, we have:
\begin{equation}
\nabla_{1_\perp}^2=\frac{1}{(1-x)^2}\nabla_{r_\perp}^2.
\tag{17}
\end{equation}
In the polar coordinate system, the Laplacian of $\nabla_{r_\perp}^2$ takes the following form:
\begin{equation}
\nabla_{r_\perp}^2=\frac{1}{r_\perp}\frac{\partial}{\partial r_\perp}(r_\perp \frac{\partial}{\partial r_\perp})+\frac{1}{r_\perp^2}\frac{\partial^2}{\partial \theta^2}.
\tag{18}
\end{equation}
Since the wave function depends only on $r_\perp$ and not on the angle $\theta$, it can ultimately be expressed as:
\begin{equation}
-\frac{\nabla_{1_\perp}^2}{x}\psi=-\frac{1}{1-x}\nabla_{\zeta}^2\psi,
\tag{19}
\end{equation}
where $\nabla_{\zeta}^2=\frac{1}{\zeta}\frac{d}{d \zeta}(\zeta \frac{d}{d \zeta})$. Then, Equation (12) reduces to:
\begin{equation}
D_\pi(Q^2)=D_1(Q^2)-\frac{1}{2}A(Q^2),
\tag{20}
\end{equation}
where
\begin{equation}
\begin{aligned}
D_1(Q^2)=&2\int_0^1dx\int\zeta d\zeta (\frac{\psi(x,\zeta)}{x(1-x)}[\frac{-1}{Q^2}(2\pi J_0(\zeta Q\sqrt{\frac{1-x}{x}})\\
&-2\pi J_0(\zeta Q\sqrt{\frac{x}{1-x}}))
(\frac{1}{1-x}\nabla_{\zeta}^2+x\ M_\pi^2)\\
&-\frac{\pi J_0(\zeta Q\sqrt{\frac{1-x}{x}})}{x}]\psi(x,\zeta)+\frac{1}{x(1-x)^2}\frac{1}{\zeta}\\
&\times\frac{\partial}{\partial \zeta}(\zeta \psi(x,\zeta) \partial_\zeta \psi(x,\zeta)))
\end{aligned}
\tag{21}
\end{equation}

Equations (16) and (20) represent the gravitational form factors in light-front QCD, and direct calculation from light-front QCD is very complicated. In the next chapter, we use the correspondence between light-front QCD and holographic QCD to compute the wave functions, which are relatively simple and effective in holographic QCD.

\section{Light-front wave function from holographic and light-front QCD correspondence}\label{sec:03}

In this chapter, we use holographic QCD to calculate the gravitational form factor $A_\pi(Q^2)$. The metric of the $AdS_5$ spacetime can be written as \cite{Karch:2006pv}:
\begin{equation}
ds^2=e^{2A(z)}(\eta_{\mu\nu}dx^{\mu}dx^{\nu}-dz^2),
\tag{22}
\end{equation}
where $A(z)=-\ln(z)$ and $z$ represents the holographic coordinate, with the UV boundary at $z=0$, and $\eta_{\mu\nu}$ (signature (+,-,-,-)) is the four-dimensional Minkowski metric.

The five-dimensional action describing the pion takes the form \cite{Brodsky:2008pf,Brodsky:2007hb,deTeramond:2013it,MartinContreras:2022xzg,deTeramond:2021lxc,Brodsky:2014yha}:
\begin{equation}
S_\pi=\int d^4xdz\sqrt{g}[g^{mn}\partial_m\Phi^*\partial_n\Phi-m_5^2\Phi^*\Phi],
\tag{23}
\end{equation}
where $\Phi$ denotes the pion field and $m_5$ is its five-dimensional mass. The equation of motion following from action (23) is:
\begin{equation}
\partial_m[\sqrt{g}g^{mn}\partial_n\Phi]-\sqrt{g}m_5^2\Phi=0,
\tag{24}
\end{equation}
with $g^{mn}=e^{-2A(z)}\eta^{mn}$. Thus, Equation (24) can be written as:
\begin{equation}
\partial_m[e^{3A(z)}\eta^{mn}\partial_n\Phi]-e^{5A(z)}m_5^2\Phi=0.
\tag{25}
\end{equation}

By defining $\omega(z)=-3A(z)$, the following result is obtained
\begin{equation}
\partial_m[e^{-\omega(z)}\eta^{mn}\partial_n\Phi]-e^{\frac{-5\omega(z)}{3}}m_5^2\Phi=0.
\tag{26}
\end{equation}

We assume that the amplitude depends solely on the holographic coordinate $z$ and propagates in the transverse space $x^\mu$ with momentum $p^\mu$. To this end, we adopt a plane wave ansatz
\begin{equation}
\Phi(x^\mu,z)=\Psi(z) e^{ip_\mu x^\mu},
\tag{27}
\end{equation}
here $p^2=p^\mu p_\mu=M_n^2$. Introducing the redefinition $\Psi(z)=e^{\frac{\omega(z)}{2}}\varphi(z)$ and performing some algebra, we obtain the Schrödinger-like equation:
\begin{equation}
-\varphi''(z)+[\frac{\omega'(z)^2}{4}-\frac{\omega''(z)}{2}+e^{\frac{-2\omega(z)}{3}}m_5^2]\varphi(z)=M_n^2\varphi(z),
\tag{28}
\end{equation}
where $M_n$ are the four-dimensional pion masses, with $n=1$ for the ground state and $n=2,3,...$ for the excited states.

The gauge-invariant definition of the energy-momentum tensor gives the following expression for the AdS matrix elements, which describe how matter fields couple to an arbitrary external source at the asymptotic boundary:
\begin{equation}
T^{mn}(x,z)=\frac{-2}{\sqrt{-g}}\frac{\delta S_\pi}{\delta g_{mn}}.
\tag{29}
\end{equation}
To obtain the explicit form of the transition amplitudes, we perturb the metric slightly away from its $AdS_5$ background. With the perturbation $\eta'_{\mu\nu}=\eta_{\mu\nu}+h_{\mu\nu}$, the interaction takes the form:
\begin{equation}
S_{int}=\frac{1}{2}\int d^4xdz\sqrt{g}h_{\mu\nu}T^{\mu\nu}+ \mathrm{o}(h^2).
\tag{30}
\end{equation}
For the graviton, perturbing the metric as $\eta'_{\mu\nu}=\eta_{\mu\nu}+h_{\mu\nu}$ in $S_G$ yields its action:
\begin{equation}
\begin{aligned}
S_h=\frac{1}{4\kappa^2}\int d^4xdz\sqrt{g}(\partial_\lambda h^{\mu\nu}\partial^\lambda h_{\mu\nu}
-\frac{1}{2}\partial_\lambda h\partial^\lambda h)+ \mathrm{o}(h^2),
\end{aligned}
\tag{31}
\end{equation}
where $\kappa$ represents the Newton constant and $h$ is the trace $h^m_m$. To arrive at (31), we have exploited the gauge transformation $h'_{\ell m} = h_{\ell m} + \partial_\ell \epsilon_m + \partial_m \epsilon_\ell$, to adopt the harmonic gauge condition $\partial_\ell h_m^\ell = \frac{1}{2} \partial_m h$.

The following integral arises from the interaction term (30), which describes how the pion mode couples to an external graviton field in AdS space:
\begin{equation}
\int d^4 x \, dz \, \sqrt{g} \, h_{\ell m} \left( \partial^\ell \Phi_{p'}^* \partial^m \Phi_p + \partial^m \Phi_{p'} \partial^\ell \Phi_p \right),
\tag{32}
\end{equation}
This expression gives the energy-momentum tensor matrix element for the hadronic transition \( p \to p' \).

The gauge invariance of $\Theta^{\ell m}$ allows us to impose the harmonic-traceless gauge $\partial_{\ell} h_{m}^{\ell} = \frac{1}{2} \partial_{m} h = 0$ and set $h_{zz} = h_{z\mu} = 0$ for the AdS graviton probe. Under these conditions, the linearized Einstein equations simplify to:
\begin{equation}
[\partial_z^2+3A'(z)\partial_z-Q^2]H(Q^2,z)=0,
\tag{33}
\end{equation}
where $h_{\mu\nu}=\epsilon_{\mu\nu}e^{-iq_Gx}H(Q^2,z), q_G^2=-Q^2$. The following boundary conditions are imposed:
\begin{equation}
H(Q^2,0)=H(0,z)=1,\ \ \partial_zH(Q^2,\infty)=0.
\tag{34}
\end{equation}

Combining Equation (4) and Equation (32), we can obtain the gravitational form factor $A(Q^2)$ as
\begin{equation}
A_\pi(Q^2)=\int dz\varphi(z)H(Q^2,z)\varphi(z).
\tag{35}
\end{equation}
The solution $H(Q^2,z)$ of Equation (33) can be written in integral form as:
\begin{equation}
H(Q^2,z)=2\int_0^1 dx x J_0(zQ\sqrt{\frac{1-x}{x}}).
\tag{36}
\end{equation}

Comparing Equations (16), (35), and (36), we can obtain \cite{Brodsky:2008pf}
\begin{equation}
\psi(x,\zeta)=\frac{1}{\sqrt{2\pi}}\sqrt{x(1-x)}\frac{\varphi(\zeta)}{\sqrt{\zeta}}.
\tag{37}
\end{equation}
where $\zeta=z$. Equation 37 is an exact form obtained through comparison, but the integral form of $H(Q^2,z)$ is not unique; it can also be written in a parametric form in $x$:
\begin{equation}
H(Q^2,z)=2\int_0^1 dxw'(x) w(x) J_0(zQ\sqrt{\frac{1-w(x)}{w(x)}}),
\tag{38}
\end{equation}
where $w(0)=1, w(1)=1$. This means that Eq. 37 is not unique; additional factors can be introduced. We assume its form to be
\begin{equation}
\psi(x,z)=\frac{1}{\sqrt{2\pi}}\sqrt{x(1-x)}\frac{\varphi(z)}{\sqrt{z}}f(x,z),
\tag{39}
\end{equation}
where $f(x,z)$ is an unknown function.

In holographic QCD, using a non-conformal model, one can obtain the pion wave function, which incorporates non-perturbative physical effects. Here we set $f(x,z) = f(x)$, meaning that the dependence on $z$ is absorbed into the five-dimensional pion wavefunction $\overline{\varphi}(z)$, which is obtained from the non-conformal holographic QCD model described in Sec. IV. The wave function then takes the form:
\begin{equation}
\psi(x,z)=\frac{1}{\sqrt{2\pi}}\sqrt{x(1-x)}\frac{\overline{\varphi}(z)}{\sqrt{z}}f(x).
\tag{40}
\end{equation}
Substituting Equation (40) into Equation (16), we obtain:
\begin{equation}
A_\pi(Q^2)=2\int_0^1dx\int_0^\infty dz\overline{\varphi}^2(z) x(J_0(z Q\sqrt{\frac{1-x}{x}})f^2(x)).
\tag{41}
\end{equation}
Furthermore, the gravitational form factor can also be written as the first moment of the generalized parton distribution function (GPD):
\begin{equation}
A_\pi(Q^2)+\xi^2D_\pi(Q^2)=\int_0^1dx xH_\pi(x,Q^2,\xi),
\tag{42}
\end{equation}
where $H_\pi(x,Q^2,\xi)$ is the GPD of the pion, with $\xi$ denoting the skewness. In the forward limit: $Q^2\rightarrow 0, \xi \rightarrow 0$, the generalized parton distribution $H_\pi(x,Q^2,\xi)$ reduces to the parton distribution function, whose asymptotic form behaves as $(1-x)^2$ as $x\rightarrow 1$, in agreement with perturbative theory and other models \cite{Yuan:2003fs,Brodsky:2006hj,Hecht:2000xa,Aicher:2010cb,Chang:2014lva,Chen:2016sno}. From Equation (42), we obtain that the asymptotic behavior of the integral kernel of \(A_\pi(0)\) is \((1-x)^2\) as \(x \to 1\). Comparing Equations (41) and (42), we find that \(f(x) \to 1-x\) as \(x \to 1\). Furthermore, our previous derivation has assumed that the two-parton distribution is symmetric, i.e., the wave function is invariant under the exchange of $x$ and $1-x$. Based on the above discussion, we choose the form of $f(x)$:
\begin{equation}
f(x)=\beta x(1-x),
\tag{43}
\end{equation}
where the coefficient $\beta=\sqrt{30}$ is fixed by the normalization condition $A_\pi(0)=1$.

In light-front QCD, the light-front wave function renormalization condition is given by \cite{Brodsky:2014yha}
\begin{equation}
\int^1_0 dx\int d^2b_\perp |\psi(x,b_\perp)|^2=1,
\tag{44}
\end{equation}
where $\xi^2=x(1-x)b_\perp^2$. Substituting our light-front wave function into Equation (44), our result also satisfies the renormalization condition.

The dual form of Equation (39) is derived from comparing the gravitational form factors in light-front QCD and holographic QCD, using a conformal metric that does not incorporate QCD features such as confinement. Equations (40) and (43) together constitute our effective wave function, obtained employing a non-conformal metric and the asymptotic form of pion GPD, in order to phenomenologically account for confinement and higher-order effects. Since the strongly correlated configurations correspond to the intermediate $x$ region, this effective wave function leads to a stronger suppression at the endpoints, which in turn increases the contributions from the strongly correlated regions. Although our assumption in Equation (43) is consistent with some physical features, it is not rigorously derived but serves as an effective model. In the next chapter, we will use this wave function to calculate the gravitational form factor $A(Q^2)$ of the pion to further examine our model, and then proceed to compute the gravitational form factor $D(Q^2)$ as well as the radius of the pion.

\section{Results}\label{sec:04}

In this chapter, we apply this wave function to compute various physical quantities of the pion, including its gravitational form factors and radius. The standard AdS/CFT duality pertains to supersymmetric Yang–Mills theory. In order to describe QCD, one must break conformal invariance and reproduce the key properties of QCD, such as confinement and asymptotic freedom—this constitutes holographic QCD. To this end, we introduce three different holographic models and use them to compute the five-dimensional pion wave function, which is then substituted into Equation (40) to obtain the light-front wave function.

The soft-wall holographic model is widely preferred for two main reasons: its smooth infrared cutoff is more natural than the abrupt hard-wall cutoff, and it successfully reproduces linear Regge behavior. The original soft-wall metric is given by \cite{Branz:2010ub}:
\begin{equation}
ds^2=\frac{1}{z^2}(\eta_{\mu\nu}dx^{\mu}dx^{\nu}+dz^2),
\tag{45}
\end{equation}
The action is given by:
\begin{equation}
S=\int d^4xdz\sqrt{-g}e^{\alpha_0(z)}\mathcal{L}.
\tag{46}
\end{equation}
Here $\mathcal{L}$ is the Lagrangian density, and the dilaton field $\alpha_0(z)=c_0z^2$ (with $c_0$ a free parameter). This action captures the confinement properties of QCD. 

The second model is based on a modified $AdS_5$ metric. In contrast to the soft-wall model, where a dilaton field is introduced in the action, this model introduces a warp factor directly into the metric. The metric is given by:
\begin{equation}
ds^2=\frac{e^{\alpha_1(z)}}{z^2}(\eta_{\mu\nu}dx^{\mu}dx^{\nu}+dz^2),
\tag{47}
\end{equation}
where $\alpha_1(z)=c_1z^2$ (with $c_1$ a free parameter). This model encodes the confinement effects of QCD into the background metric rather than into the action, and has been used to study hadronic structure \cite{FolcoCapossoli:2019imm,FolcoCapossoli:2020pks}.

The third model is based on our previous work \cite{Deng:2026suw}, which also modifies the $AdS_5$ metric. The metric for this model is given by:
\begin{equation}
\begin{aligned}
ds^2&=\frac{e^{2k_{1}^{2}z^{2}(1-k_{2}tanh(k_{1}^{2}z^{2}))}}{z^2}(\eta_{\mu\nu}dx^{\mu}dx^{\nu}+dz^2)\\
&=e^{2A(z)}(\eta_{\mu\nu}dx^\mu dx^\nu+dz^2),
\end{aligned}
\tag{48}
\end{equation}
where $k_1$ and $k_2$ denote free parameters, and the warp factor $A(z)$ takes the form:
\begin{equation}
A(z)=-\mathrm{ln}(z)+\rho(z),\ \ \rho(z)=k_{1}^{2}z^{2}(1-k_{2}tanh(k_{1}^{2}z^{2})).
\tag{49}
\end{equation}
The UV boundary condition of the $\rho(z)$ field can be explicitly written as:
\begin{equation}
\rho(z)=k_{1}^{2}z^{2}-k_2k_{1}^{4}z^{4}.
\tag{50}
\end{equation}
The dimension-4 gauge-invariant gluon condensate of the boundary QCD is holographically encoded in the coefficient of the $z^4$ term in the UV expansion of the dilaton. In the infrared regime, the dilaton instead behaves as:
\begin{equation}
\rho(z)=k_{1}^{2}z^{2}-k_2k_{1}^{2}z^{2}.
\tag{51}
\end{equation}
Consequently, the UV component that holographically represents the gluon condensate re-emerges in the IR as a quadratic term, thus contributing to the infrared dynamics.

As shown in Fig.~1, we compute the five-dimensional pion wave functions from the three holographic models. The parameter $c_0\approx 0.40\ \text{GeV}^2$ in model~1 (blue) and the parameter $c_1\approx-0.17\ \text{GeV}^2$ in model~2 (purple) are both determined by the lattice data point $A_\pi(Q^2=0.07\ \text{GeV}^2)\approx0.96$ (the first lattice data point). Model~3 has two parameters, $k_1=0.22\ \text{GeV}$ and $k_2=3.799$, which are determined by $m_\pi=0.17\ \text{GeV}$ and the same lattice data point $A_\pi(Q^2=0.07\ \text{GeV}^2)\approx0.96$ (black).
\begin{figure}
	\centering
	\includegraphics[width=7.5cm]{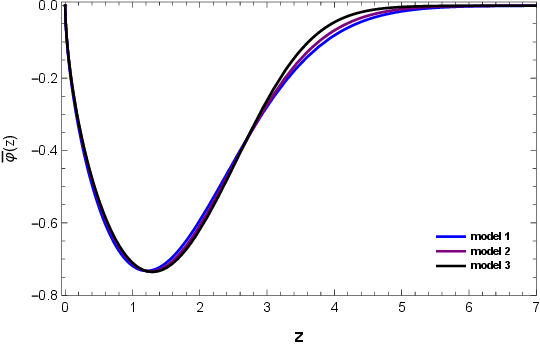}
	\caption{\label{figure 1}Comparison of the five-dimensional wave function of the pion obtained from the three holographic QCD models.}
\end{figure}

Substituting the calculated five-dimensional wave functions into Equation (40) and combining with Equation (43), we obtain the gravitational form factor $A(Q^2)$ of the pion.
\begin{figure}
	\centering
	\includegraphics[width=7.5cm]{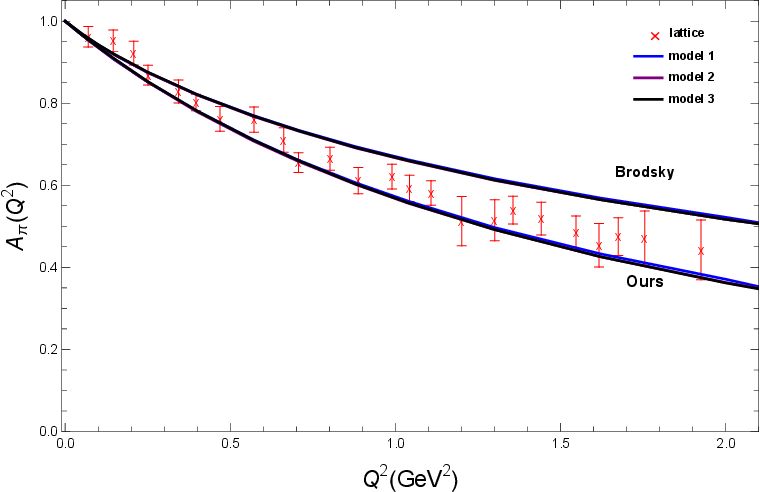}
	\caption{\label{figure 2}Gravitational form factor $A(Q^2)$ of the pion. The red crosses represent lattice QCD results at $m_\pi=0.17\ \text{GeV}$ and scale $\mu^2=4\ \text{GeV}^2$ \cite{Hackett:2023nkr}. The solid blue, purple, and black lines denote the results obtained from holographic models 1, 2, and 3, respectively. The results denoted as "Brodsky" are obtained from the original light-front wave function [Eq. (37)] \cite{Brodsky:2008pf}, and those denoted as "Our" are obtained from our proposed form [Eq. (40)].}
\end{figure}
As shown in Fig.~2, we compare the gravitational form factor $A_\pi(Q^2)$ of the pion calculated by two forms of light-front wave functions under three different holographic models. The results obtained from the three holographic approaches are all very close to each other, and the results calculated with our light-front wave function are closer to the lattice QCD data. The $\chi^2/\mathrm{dof}$ values for models 1, 2, and 3 are 1.11, 1.31, and 1.24, respectively, where $\chi^2 = \sum_i (A_i^{\text{our}} - A_i^{\text{lattice}})^2 / \sigma_i^2$ and $\sigma_i$ are the lattice errors, whereas the $\chi^2/\mathrm{dof}$ for the Brodsky model is about 3. These results indicate that the models provide a reasonably good description of the lattice data and demonstrate that our adopted form of the light-front wave function provides a better description of the gravitational structure than the original one. Moreover, substituting the original form Eq. (37) into Eq. (20) leads to a divergence, making it impossible to compute the gravitational form factor $D(Q^2)$. Therefore, in the following, we employ our light-front wave function to calculate the gravitational form factor D and use model 3 to investigate its dependence on the pion mass.

Using the computed five-dimensional wave function together with Equations (20), (40), and (43), we obtain the gravitational form factor $D_\pi(Q^2)$.
\begin{figure}
	\centering
	\includegraphics[width=7.5cm]{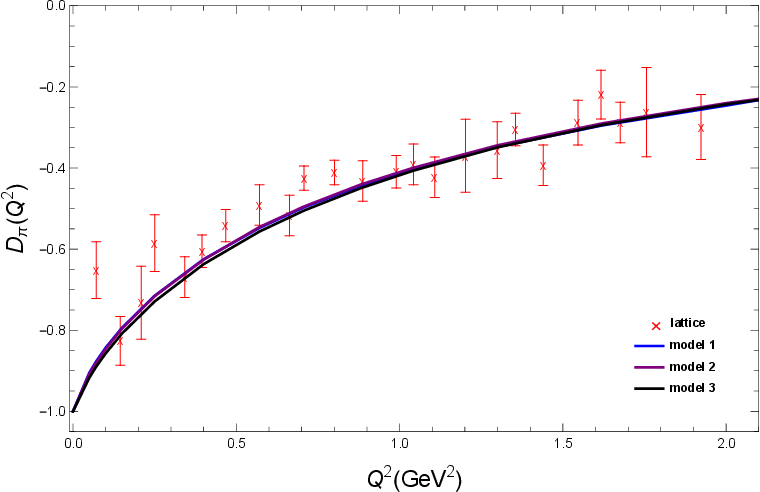}
	\caption{\label{figure 2}Gravitational form factor $D(Q^2)$ of the pion. The red crosses represent lattice QCD results at $m_\pi=0.17\ \text{GeV}$ and scale $\mu^2=4\ \text{GeV}^2$ \cite{Hackett:2023nkr}. The solid blue, purple, and black lines denote the results obtained from holographic models 1, 2, and 3, respectively.}
\end{figure}
As shown in Fig.~3, the results calculated from the three holographic models are close to each other. The $\chi^2/\mathrm{dof.}$ values for models 1, 2, and 3 are 1.32, 1.31, and 1.56, respectively. These results are overall consistent with the lattice data. The analysis shows that the deviations are mainly concentrated in a few points at the low-$x$ end , while the remaining lattice points show good agreement.

\begin{figure}
	\centering
	\includegraphics[width=7.5cm]{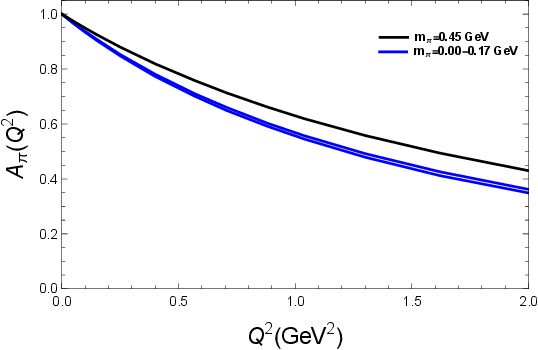}
	\caption{\label{figure 2}Gravitational form factor $A(Q^2)$ of the pion at different masses. The blue and black lines represent the results from Model~3 for $m_\pi = 0$--$0.17\ \text{GeV}$ and $m_\pi = 0.45\ \text{GeV}$, respectively.}
\end{figure}

As shown in Figure 4, we show the gravitational form factor $A_\pi(Q^2)$ at different masses obtained from Model~3. The blue curve The blue curve corresponds to the pion mass range $m_\pi=0-0.17\ \mathrm{GeV}$ (with $k_2=3.534-3.799$), where mass is varied by adjusting $k_2$ and the black line corresponds to the pion mass $0.45\ \mathrm{GeV}$. We find that the gravitational form factor $A_\pi(Q^2)$ increases with increasing mass. In addition, near the physical pion mass, the gravitational form factor exhibits only a mild variation.
\begin{figure}
	\centering
	\includegraphics[width=7.5cm]{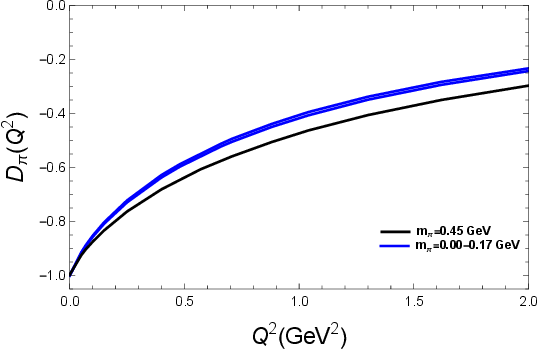}
	\caption{\label{figure 2}Gravitational form factor $D(Q^2)$ of the pion at different masses. The blue and black lines represent the results from Model~3 for $m_\pi = 0$--$0.17\ \text{GeV}$ and $m_\pi = 0.45\ \text{GeV}$, respectively.}
\end{figure}
As shown in Figure 5, we show the gravitational form factor $D_\pi(Q^2)$ at different masses obtained from Model~3. The blue curve corresponds to the pion mass range $m_\pi=0-0.17\ \mathrm{GeV}$ and the black line corresponds to the pion mass $0.45\ \mathrm{GeV}$. We find that the gravitational form factor $D_\pi(Q^2)$ decreases with increasing mass. In addition, near the physical pion mass, the gravitational form factor exhibits only a mild variation. The gravitational form factor $D_\pi(Q^2)$ is found to be largely insensitive to the pion mass, confirming that strong interaction dynamics dominate the mechanical structure of the pion. 

As shown in Figs.~2 and 3, the gravitational form factors $A(Q^2)$ and $D(Q^2)$ obtained from the three holographic models are mutually consistent. This indicates that light-front holographic QCD duality approach is robust and exhibits only a mild dependence on the specific holographic model within the class of models considered here.

The internal structure of the pion is characterized by its radii, which provide a window into nonperturbative QCD. These radii are extracted from various physical observables, each revealing a different facet of the pion's interior. Beyond their structural significance, they also serve as critical inputs for precision tests of the Standard Model. Notably, accurate measurements of pion radii impose tight constraints on theoretical frameworks, including lattice QCD, Dyson-Schwinger equations, and light-front holographic QCD.

The radii are the key parameters by which GFFs A and D are characterized:
\begin{equation}
\langle r_{A}^2\rangle=-\frac{6}{A(0)}\frac{dA(Q^2)}{dQ^2},
\tag{52}
\end{equation}
\begin{equation}
\langle r_{D}^2\rangle=-\frac{6}{D(0)}\frac{dD(Q^2)}{dQ^2}.
\tag{53}
\end{equation}
Our results are $r_{A}\approx0.39$\ fm and $r_{D}\approx0.95$\ fm. These values approximate the corresponding radii $r_{A}= 0.32-0.39$\ fm and $r_{D}= 0.82-0.88$\ fm  obtained from two-photon scattering \cite{Kumano:2017lhr}.

In this chapter, we have computed the pion gravitational form factors within the light-front holographic QCD framework. Starting from the five-dimensional pion wave function, which encodes essential QCD features such as confinement, we derived an effective light-front wave function via the holographic duality and used it in the light-front QCD formula. 
This effective wave function imposes a stronger suppression in the endpoint regions, thus boosting the contributions from the strongly correlated regions. We compare three different holographic models and find that the resulting gravitational form factors are close to each other and are in good agreement with the lattice data.

\section{Conclusions and discussion}\label{sec:05}

In this study, we constructed an effective light-front wave function (LFWF) for the pion by integrating principles from light-front QCD and holographic QCD duality, which incorporates the asymptotic behavior of the pion’s parton distribution function as $x \to 1$, as well as parton symmetry considerations. Using this wave function, we calculated the gravitational form factors $A(Q^2)$ and $D(Q^2)$ of the pion. Notably, our model addresses a limitation of previous approaches by enabling the calculation of $D(Q^2)$, which was divergent in the original formulation, and yields results that are consistent with lattice QCD calculations. 

Specifically, we derived the gravitational form factor $A(Q^2)$ from light-front QCD and the AdS/CFT correspondence, from which we obtained the conformal light-front wave function. To more closely approximate real QCD, we modified this wave function by incorporating a function $f(x, z)$. By absorbing the $z$-dependence into the five-dimensional pion wave function obtained from holographic QCD with a non-conformal background, we set $f(x, z) = f(x)$. The asymptotic form $f(x) \sim 1-x$ was then determined by equating the forward limit of the gravitational form factor with the first moment of the generalized parton distribution and combining this with its asymptotic form as $x \to 1$. Considering the symmetry of the wave function under the exchange $x \leftrightarrow 1-x$, we adopted the form $f(x) =\beta x(1-x)$. This factor suppresses endpoint contributions—where the two partons are spatially far apart—thereby effectively enhancing the short-range parton-parton correlations. With this input, we obtained an effective light-front wave function and used it to compute the pion gravitational form factors

While our results show good agreement with lattice QCD data, several limitations of this study must be acknowledged. First, the light-front wave function presented here is not derived strictly from first principles but is instead constructed based on physical considerations, making it an effective wave function. Second, our current model does not incorporate contributions from higher Fock states. These limitations point to important directions for future research. Future work should aim to develop more comprehensive model studies that include higher-order effects, rigorously derive the form of the wave function from first principles, and extend this investigation to the internal structure of other hadrons.

In summary, our results demonstrate that the correspondence between light-front QCD and holographic QCD offers a distinct and useful perspective for studying hadron structure. Despite the acknowledged limitations, this model provides important insights for future research on the internal structure of hadrons and other non-perturbative phenomena in strong interactions.

\section*{Acknowledgments}

We thank Hai-cang Ren and  Yan-Qing Zhao for useful discussions. This work is supported in part by the National Key Research and Development Program of China under Contract No. 2022YFA1604900. This work is also partly supported by the National Natural Science Foundation of China(NSFC) under Grants No. 12435009, and No. 12275104.

\section*{References}

\bibliography{ref}

\end{document}